# From Prompt to Embodied Simulation: Using Generative AI to Create AR Physics Learning Tools

*Ofek Levy,[1] Joshua Glazer,[1] Noah D. Finkelstein,[2] and Yossi Ben-Zion[1,*]*

*[1]Department of Physics, Bar-Ilan University, Ramat Gan IL52900, Israel*

*[2]Department of Physics, University of Colorado Boulder, Boulder, Colorado 80309, USA*

**benzioy@biu.ac.il*

Spread your thumb and index finger in the air, and a virtual lamp in the room changes color. Computer simulations have a long and well-documented record of supporting physics learning.[1,2] They can support the understanding of abstract physical concepts by making them interactive and by inviting students to play with parameters and explore them. In this paper, we show how a structured natural-language prompt can generate a browser-based, hand-controlled augmented-reality (AR) physics simulation, and we describe its use in an introductory physics class.

Recent work has shown that generative artificial intelligence is changing who gets to build such tools: using a reusable prompt structure, a teacher or a student with no coding background can ask a large language model (LLM) to generate a working, browser-based simulation, and then refine it in natural language.[3,4]

Those simulations, however, share a common interface: the learner sits in front of a screen and changes the parameters by dragging sliders, or otherwise inputting values. A slider is a powerful and familiar control, but it is an abstraction: a labeled bar that one reads and adjusts. Research in the learning sciences has shown that in some cases there is an advantage to learning through sensorimotor experience: students who physically experience a phenomenon can learn it better than students who only watch.[5] Embodied cognition and learning have a rich history across educational environments and may (and certainly ought to) be something that distinguishes human cognition from AI.[6] Gesture plays a central role in human communication and reasoning, even though research on student thinking often relies heavily on language and mathematical representations.[7] If amplitude and wavelength are represented spatially in a wave diagram, then perhaps the most natural control for them is not a slider or input box, but the hand itself moving in space.

Building on earlier prompt-generated browser simulations, we extend the approach to embodied, camera-based interaction. The new principle is simple: a live camera sees the learner's hands, and the gestures captured by the camera control the parameters of the problem. This also brings challenges that slider simulations do not have: recognizing the gesture reliably, keeping the motion stable, aligning the drawing with the mirrored video, and behaving sensibly when the hand is lost. Until recently, building such a simulation, which combines hand tracking, real-time graphics, and a physical model, demanded a rare combination of skills. Today it can often be produced in a common LLM tool, and embodied AR is within reach of many teachers and students.

We present the approach through a demonstration: a sine wave and a lamp, drawn over a live view of the user's environment, both controlled by the same motion of the hand (Fig. 1). The gesture is familiar to most students from touchscreen interactions: the pinch-and-spread motion of the thumb and index finger. Here, however, the pinch does not zoom a picture but tunes a physical quantity: opening the fingers vertically raises the amplitude and strengthens the lamp's light, and opening them horizontally lengthens the wavelength and shifts the color toward the red end of the visible spectrum. The wave is the mathematical representation, and the virtual lamp provides an everyday visual interpretation: two separate objects obeying the same two parameters. The mapping is qualitative and meant for teaching: the drawn height stands for amplitude, the glow for intensity, and the color for the order of the visible wavelengths; the simulation is not a quantitative model of an electromagnetic field.

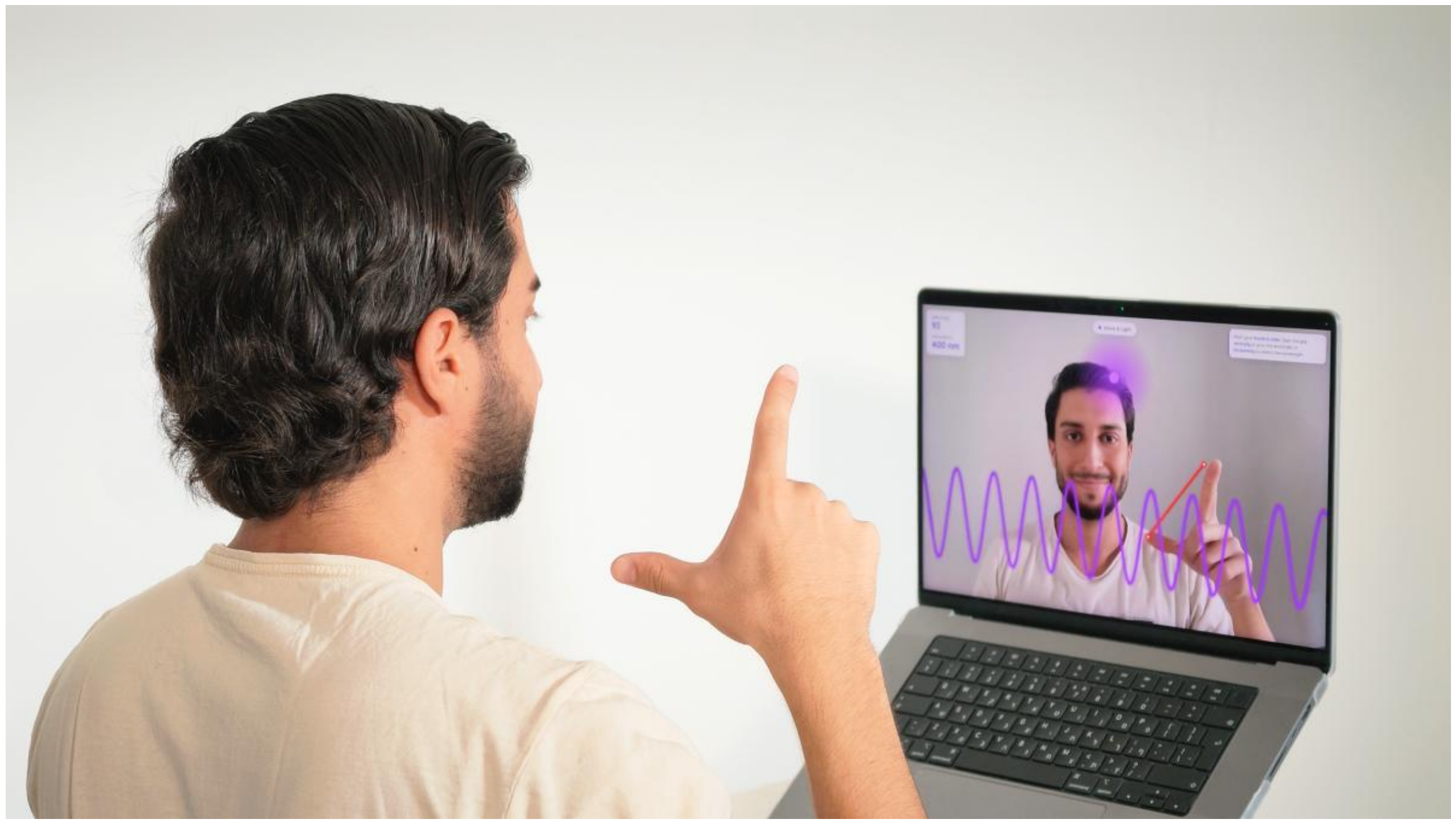

**Fig. 1.** The wave-and-lamp simulation. The gap between the thumb and index finger controls the amplitude and the wavelength of the wave; the lamp's glow and color follow the same two parameters. The current values are shown on the screen.

## The prompt

Each simulation we present runs as a single HTML file in an ordinary browser, with no installation and no special hardware beyond a camera. Two core components are needed. The first is MediaPipe Hands, a ready-made library that detects the finger joints in the camera image in real time. The second is a transparent drawing layer above the video, on which the physics is drawn. When the drawing is three-dimensional, we also use the Three.js library.

The prompt has four elements:

- **The tools:** which libraries are used and how the camera is shown.
- **Display:** which objects are drawn over the camera image and what each of them represents.

- **Hand controls:** which finger motion controls which quantity, and what the simulation ignores.
- **Optimization:** requirements that prevent jumps and jitter.

The elements are fixed; only their content changes from one simulation to the next.

We start with Gemini.[8] Two things are done in the interface:

1. Choose the latest Pro model available; at the time of writing, this is Gemini 3.1 Pro.

2. Switch to Canvas mode: a mode in which the model writes the code and runs it immediately in the same window, so the simulation is seen without leaving the conversation.

Paste the following prompt, built from the four elements, one paragraph per element:

> *You are a designer of physics simulations for teaching. Create an interactive augmented-reality simulation using MediaPipe Hands. Show the live camera feed full screen, mirror it horizontally like a mirror, and track one hand.*
>
> *Draw two elements: a glowing lamp at the top and a sine wave at the bottom. They are not physically connected; both respond to the same two parameters. The sine wave is the mathematical representation, and the lamp's light is the physical manifestation. Display the current values of the amplitude and the wavelength in the bottom-right corner.*
>
> *The simulation has exactly one controller: the gap between the tip of the thumb and the tip of the index finger. When the gap is vertical, its size sets the amplitude: a wide gap means a tall wave and a bright lamp, and a narrow gap means a low wave and a dim lamp. When the gap is horizontal, its size sets the wavelength: a wide gap means a long wavelength and a red color, and a narrow gap means a short wavelength and a violet color, with smooth transitions between them. The wave and the lamp always share the same color. Only one parameter changes at a time, the one matching the gap's direction, while the other keeps its value; when neither direction clearly dominates, keep the previous values. Measure the gap relative to the size of the hand itself, for example the width of the palm,* to reduce sensitivity to changes in the hand's distance from the camera.
>
> *Animate the wave so that it travels at a constant speed. When the wavelength changes, keep the wave's phase continuous and the motion smooth, with no visible jumps. If the hand is lost or the tracking becomes uncertain for a moment, keep the last stable values. Make sure the drawing uses the same mirrored screen coordinates as the video, so that it matches the hand as it appears on the screen.*

In other tools, such as ChatGPT[9] and Claude,[10] the model writes the same code, but the built-in preview window may not have access to the camera. In that case, three steps:

1. Add an opening sentence to the prompt: *write the code as one complete, runnable HTML file, in a code block*, and after it paste the entire prompt.

2. Copy the resulting code into a file and save it with an .html extension.

3. Double-click the file to open it in the browser.

On the first run, the browser asks for camera access. In class, it also helps to keep the hand at a roughly constant distance from the camera, where the tracking is most stable.

## Iterations and validation

A first prompt is almost never perfect, and in AR simulations this is especially visible: they are sensitive to edge cases that appear only when a real hand moves in front of a real camera. The iterative prompt-and-refinement process follows the approach described in previous work,[3,4] and it is where the AR-specific challenges noted in the introduction are met and resolved. The loop itself is short: run, notice what behaves incorrectly, describe the problem in plain language, and send a correction prompt. After a few iterations, the model can often produce the desired behavior. Two of the sentences in the prompt above were born exactly this way.

The rule that the gap is measured relative to the size of the hand was added after the values drifted whenever the hand moved closer to or farther from the camera. The phase-continuity requirement was added after the wave shook with every change of the wavelength. In both cases the fix was a single sentence, for example: *When the wavelength changes, the wave shakes and jumps. Keep the wave's phase continuous, so the change looks like a smooth stretch rather than a jump.*

We check the result on three levels. A technical check: the tracking finds the hand in ordinary classroom lighting, the drawing moves in the same on-screen direction as the hand in the mirrored video, and the motion is smooth. A physical check: the relations between the quantities are correct, for example a high frequency goes with a short wavelength and a violet color, consistent with the ordering of colors in the visible spectrum. A pedagogical check: the gesture is natural, and operating the interface does not compete with the physics for attention.

## A flexible starting point for the teacher's creativity

The same four elements produce very different simulations, and in all of them the physics appears in the learner's own environment. In one simulation [Fig. 2(a)] the tip of the index finger is an electric charge, and a three-dimensional grid of arrows, showing the electric field of the charge, fills the space around the student at several depths in front of the camera. Each arrow points away from the charge, and the arrows are long near the finger and grow shorter farther away: Coulomb's law spread out inside the room itself. When the student moves their hand, the whole field moves with it. In a second simulation [Fig. 2(b)], each index finger is a charge, positive in one hand and negative in the other, and the arrows show their combined field: bringing the hands closer concentrates the field between them and moving them apart stretches it across the environment. In a third simulation [Fig. 2(c)], the right-hand rule is drawn on the student's own hand: three arrows, force, velocity, and field, anchored to three fingers and rotating with the hand. The full prompts are given in the appendix. The physics changes while the structure stays.

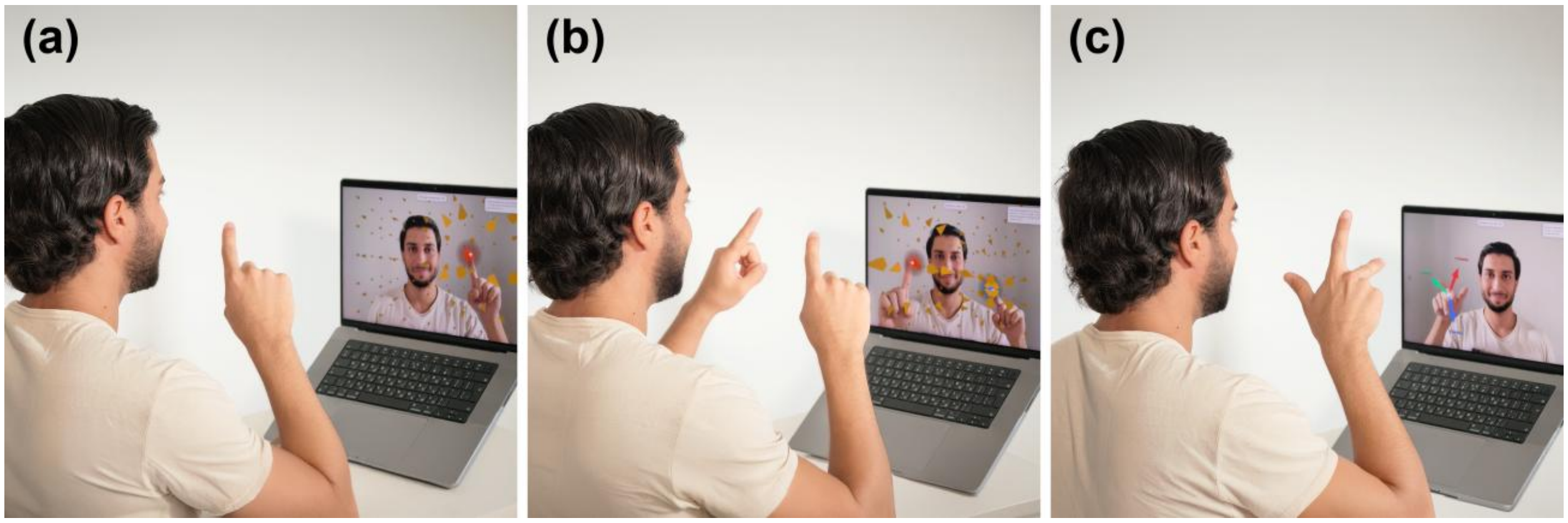

**Fig. 2.** Three simulations built from the same four-element prompt: (a) a positive charge on the fingertip, with the field arrows filling the space around the student; (b) two charges, one in each hand, with the combined field between them; (c) the right-hand rule for $F = qv \times B$, drawn on the student's own hand.

## In the classroom

We pilot-tested the wave-and-lamp simulation in a class of 29 students in an introductory radiation physics course in 2026. The students were second-year medical imaging majors. They received the initial prompt presented above, generated the simulation with it, and performed a technical and a physical validation of the result; they then explored it freely. At the end of the activity they completed an attitude survey on a 1-to-5 scale and three open questions. The survey measured student perceptions and engagement, not understanding or achievement; responses were collected with the students' consent and are reported anonymously. The survey items and the quotations below are translated from Hebrew. All 29 students agreed that the pinch-and-spread motion of the fingers helped them "feel" what wavelength is (mean 4.52). The statement about feeling the amplitude received the highest mean of all, 4.59. Controlling the wave in the air felt natural to 93% of them, and 86% reported feeling more engaged and focused than in regular learning. The two statements about the lamp also scored high: the link between color and mathematics (mean 4.45) and between the strength of the glow and the amplitude (4.48). The full survey contained nine statements, including three about a variant of the simulation in which the same parameters drive a tone instead of a lamp. All nine items received positive average ratings, with agreement ranging from 79% to 100%; we report here those that concern the light version presented in this paper.

Three themes repeat in the open answers. The first is feeling the wave: "It was easier to understand when I changed the wave myself, because I immediately saw what happened," and another student wrote: "I felt I could understand the material physically, that it entered my mind through using my own body to understand it." The second is the learning curve of the hand tracking: getting the camera to recognize the gesture reliably takes a few tries. As one student wrote: "It was harder because of the camera's recognition, but that made the simulation more of an experience. Once I understood how to work with it, I went back to it several times, and each time I noticed something new." Indeed, two students disagreed that controlling the wave in the air felt natural, and another student wrote that she understood the lamp and its glow only in hindsight, after a friend explained them. The third is the trace the experience leaves: "I feel that when I get to the exam, I will move my fingers in my head to picture the situation for myself."

These results are encouraging, but they were obtained on a small sample and without a comparison group. Follow-up studies can examine, in a controlled comparison, the contribution of the embodied interface to learning, and which topics are more effectively taught with an AR interface than with other tools. Such questions are now easier to examine locally because the technical barriers to producing AR interfaces have been substantially reduced.

The AR simulation we presented began with one structured prompt in plain language, with freely available tools, and it runs in a modern browser on commonly available computers with a camera. The four elements of the prompt carry over to many other phenomena. Teachers can use the prompt to design AR simulations of their own, or even guide their students to create such simulations themselves, with validation and iterations as part of the process. The rest is in the teacher’s hands.

# From Prompt to Embodied Simulation: Online Appendix

This appendix extends the paper with additional prompts built from the same four elements presented in the paper: the tools, the display, the hand controls, and the optimization requirements. The prompts are offered as flexible starting points and sources of inspiration. They were tested by the authors, may require minor iteration depending on the model, and can be adapted to different topics, levels, and classroom contexts.

## Prompt A: A single charge and its field in the room

*You are a designer of physics simulations for teaching. Create an interactive augmented-reality simulation using MediaPipe Hands and Three.js. Show the live camera feed full-screen, mirror it horizontally like a front-facing camera, and track one hand.*

*The tip of the index finger represents a positive electric charge, drawn as a glowing red sphere with a plus sign. The fingertip position determines the charge position: the student is holding the charge in their hand.*

*Overlay a perspective-rendered 3D grid of small arrows on the video at several depths in front of the camera, showing the electric field at sampled points throughout the visible space. Each arrow points radially away from the charge in three dimensions, and its length represents the field strength: arrows are longer near the charge and become progressively shorter with distance in every direction. When the student moves their finger, the entire field must update immediately and move with it. Very close to the charge, cap the arrow length so that the visualization remains readable. The arrows must move smoothly, without jitter or flicker.*

*If the hand is temporarily not visible, keep the charge at its last stable position. Apply the same horizontal reflection to the camera feed, the tracked landmarks, and the rendered overlay so that they remain aligned: when the hand moves to the right on the screen, the charge must move to the right with it.*

## Prompt B: Two charges, one in each hand

*You are a designer of physics simulations for teaching. Create an interactive augmented-reality simulation using MediaPipe Hands and Three.js. Show the live camera feed full-screen, mirror it horizontally like a front-facing camera, and track two hands.*

*Each index fingertip represents an electric charge: one is a positive charge, drawn as a glowing red sphere with a plus sign, and the other is a negative charge, drawn as a glowing blue sphere with a minus sign. The fingertip positions determine the charge positions: the student is holding the two charges in their hands.*

*Assign each detected hand to one charge and preserve that assignment across frames so that the positive and negative charges do not suddenly switch hands.*

*Overlay a perspective-rendered 3D grid of small arrows on the video at several depths in front of the camera, showing the combined electric field at sampled points throughout the visible space. Each arrow points in the direction of the net electric field in three dimensions, and its length represents the field strength. The arrows point away from the positive charge and toward the negative charge. When the fingers move closer together, the field between them becomes stronger and more concentrated; when they move farther apart, the field extends across a larger region of the scene. Very close to either charge, cap the arrow length so that the visualization remains readable. The arrows must move smoothly, without jitter, flicker, or sudden changes in direction.*

*If only one hand is visible, keep the other charge at its last stable position. Apply the same horizontal reflection to the camera feed, both sets of tracked landmarks, and the rendered overlay so that they remain aligned: when a hand moves to the right on the screen, its charge must move to the right with it.*

## Prompt C: The right-hand rule for F = qv × B, on your own hand

Unlike the other simulations, this one displays the camera feed without mirroring. A mirror image reverses one axis and therefore reverses handedness: a right hand in a mirror appears to be a left hand, and the three arrows would form a left-handed coordinate system on the screen, inverting the very rule the simulation is intended to teach.

*You are a designer of physics simulations for teaching. Create an interactive augmented-reality simulation using MediaPipe Hands and Three.js. Display the live camera feed full-screen, without horizontal mirroring, exactly as captured by the camera, and track one hand.*

*The simulation demonstrates the right-hand rule for the magnetic force on a moving positive charge, $F = qv \times B$, visualized directly on the student's hand. When the student extends the thumb, index finger, and middle finger so that they are approximately perpendicular to one another, anchor three labeled 3D arrows at the center of the palm: an arrow labeled F pointing in the direction of the thumb, an arrow labeled v pointing in the direction of the index finger, and an arrow labeled B pointing in the direction of the middle finger. Place a small glowing sphere at the point where the three arrows meet. Give each arrow a distinct, clearly distinguishable color, and attach its label to it.*

*The simulation has exactly one controller: the directions of the student's fingers. The arrows must always point in the directions indicated by the fingers and must move and rotate smoothly with the hand, without jitter. When the gesture is released, the arrows disappear; when the gesture is formed again, they reappear.*

*Do not mirror the video. Display it exactly as captured by the camera so that a right hand appears on the screen with its true handedness and the three arrows preserve the correct right-handed relationship. The arrows must remain aligned with the hand on the screen and move together with it.*